\documentclass[sigconf]{acmart}
\usepackage{bm}
\usepackage{multirow}
\usepackage{booktabs}

\AtBeginDocument{%
  }

\renewcommand\footnotetextcopyrightpermission[1]{}

\begin{document}

\title{Decision-Induced Ranking Explains Prediction Inflation and Excessive Turnover in SPO-Based Portfolio Optimization}

\author{Yi Wang}
\email{apple9238@fuji.waseda.jp}
\affiliation{%
  \institution{Waseda University}
  \city{Tokyo}
  \country{Japan}
}

\author{Takashi Hasuike}
\email{thasuike@waseda.jp}
\affiliation{%
  \institution{Waseda University}
  \city{Tokyo}
  \country{Japan}
}

\begin{abstract} Decision-focused learning (DFL) is attractive for portfolio optimization because it trains predictors according to downstream decision quality rather than prediction accuracy alone.  However, Smart Predict-then-Optimize (SPO)-based DFL may produce inflated return signals and unstable portfolio reallocations.  This study provides a KKT-based interpretation showing that portfolio decisions can be viewed as ranking over risk- and transaction-cost-adjusted marginal scores.  Empirically, we examine prediction inflation and excessive turnover in SPO-trained portfolios, and evaluate clipping, min--max rescaling, and partial portfolio adjustment as practical stabilization mechanisms.  The results suggest that realistic output constraints and portfolio-level turnover control improve the implementability of SPO-based portfolio strategies. 
\end{abstract}


\maketitle

\section{Introduction}

Portfolio optimization has been a central problem in financial decision-making since the classical mean--variance framework~\cite{markowitz1952portfolio}. 
Learning-based approaches have been widely used to support this task by estimating asset returns, risks, or other predictive signals from market data~\cite{freitas2009prediction,krauss2017deep,paiva2019decision,ma2021portfolio,behera2025optimizing}. 
Among them, predict-then-optimize (PtO) methods first train a prediction model and then use the predicted parameters as inputs to a downstream portfolio optimizer. 
This paradigm preserves the structure and interpretability of classical portfolio optimization models, and allows constraints, risk terms, and transaction-cost-related considerations to be explicitly incorporated into the decision problem.
However, financial return prediction remains difficult because asset returns are noisy, nonlinear, and non-stationary~\cite{deboeck1994trading}. 
Although modern machine learning and deep learning models have been extensively studied in financial prediction tasks, lower prediction error does not necessarily translate into better portfolio decisions. 
A small error in predicted returns may change the ranking of assets or lead the optimizer to a very different allocation. 
This mismatch motivates decision-focused learning (DFL), and in particular the Smart Predict-then-Optimize (SPO) framework, which trains prediction models according to downstream decision quality rather than prediction accuracy alone~\cite{elmachtoub2022smart,mandi2024decision,wilder2019melding}. 
Importantly, the downstream optimization model can explicitly incorporate portfolio constraints, risk preferences, and market frictions, rather than relying on a black-box model to implicitly learn these decision structures. 
This separation is particularly useful in portfolio optimization, where constraints, risk preferences, and trading frictions are often better represented explicitly in the optimizer than implicitly absorbed by a black-box prediction model.
In this paper, we use DFL to denote the broader decision-focused learning paradigm, and use SPO to refer specifically to the SPO+-surrogate-based method adopted in our experiments. 
This follows recent terminology~\cite{mandi2024decision} that treats SPO as a specific surrogate-based approach within the broader DFL framework.

\paragraph{PtO and DFL}
In a standard predict-then-optimize (PtO) framework, the predictor is trained by minimizing a prediction loss between the predicted parameter $\hat{\bm c}$ and the true parameter $\bm c$, for example
\begin{equation}
    \ell_{\mathrm{PtO}}(\hat{\bm c},\bm c)
    =
    \|\hat{\bm c}-\bm c\|_2^2 .
\end{equation}
The resulting prediction is then passed to a downstream optimization problem:
\begin{equation}
    \bm z^*(\hat{\bm c})
    \in
    \arg\min_{\bm z \in \mathcal{Z}}
    f(\bm z;\hat{\bm c}),
\end{equation}
where $\bm z$ denotes the decision variable and $\mathcal{Z}$ denotes the feasible set.

In contrast, DFL trains the predictor according to the quality of the downstream decision. 
A typical decision loss can be written as regret:
\begin{equation}
    \ell_{\mathrm{DFL}}(\hat{\bm c},\bm c)
    =
    f(\bm z^*(\hat{\bm c});\bm c)
    -
    f(\bm z^*(\bm c);\bm c),
\end{equation}
where $\bm z^*(\bm c)$ denotes the oracle decision obtained using the true parameter, and $\bm z^*(\hat{\bm c})$ denotes the decision induced by the predicted parameter. 
This distinction highlights that PtO evaluates predictions by numerical accuracy, whereas DFL evaluates predictions by the quality of the decisions they induce.

Several studies have applied decision-aware or end-to-end learning frameworks to portfolio optimization. 
Butler and Kwon~\cite{butler2023integrating} integrate regression prediction models with mean--variance optimization using a decision-aware training objective. 
Costa and Iyengar~\cite{costa2023distributionally} develop an end-to-end distributionally robust portfolio construction framework, where the prediction model is trained through a robust portfolio optimization layer. 
Anis and Kwon~\cite{anis2025end} study an end-to-end decision-based framework for cardinality-constrained portfolio optimization. 
Kim~\cite{kim2025semi} proposes a semi-decision-focused learning framework with deep ensembles for robust portfolio optimization. 
Kim et al.~\cite{kim2025estimating} apply decision-focused learning to covariance estimation for global minimum-variance portfolio construction. 
More recently, Lee et al.~\cite{lee2025return} investigate how DFL changes return prediction models in mean--variance portfolio optimization.

This study aims to clarify the behavior of SPO-based portfolio optimization from both mechanistic and empirical perspectives. 
We provide a KKT-based interpretation suggesting that the downstream portfolio optimizer can transform predicted returns into risk- and transaction-cost-adjusted marginal scores, which connects SPO-based learning to a ranking-driven allocation mechanism. 
We then examine how this mechanism appears in practice through prediction inflation and excessive turnover, and evaluate simple stabilization strategies based on prediction clipping, min--max rescaling, and partial portfolio adjustment. 
The results suggest that realistic constraints on predicted returns and portfolio-level turnover control can improve the practical implementability of SPO-based portfolio strategies.

\section{Discussion on the Mechanism of DFL in Portfolio Optimization}
\subsection{Related Work}

Lee et al.~\cite{lee2025return} provide a relatively detailed investigation of the mechanism of decision-focused learning (DFL) in portfolio optimization. They observe that DFL-trained models may lead to extremely concentrated portfolios. In their experiments, the resulting portfolios often allocate capital only to the minimum number of assets permitted by the imposed constraints, rather than producing broadly diversified allocations. This finding suggests that DFL does not simply improve return forecasts in a conventional prediction-oriented sense. Instead, it can reshape the prediction model in a way that directly favors the downstream portfolio decision, even if this leads to stronger asset selection and higher concentration.

More generally, Mandi et al.~\cite{mandi2022decision} interpret decision-focused learning through the lens of learning to rank. They argue that, in combinatorial optimization, the objective function induces a partial ordering over feasible solutions, and learning this ordering is sufficient to obtain low regret. In particular, if the predicted objective function ranks the true optimal solution ahead of the competing feasible solutions, the downstream decision remains optimal even when the predicted parameters are not numerically accurate. This perspective suggests that DFL is not merely about estimating the true cost or return vector, but about preserving the decision-relevant ordering induced by the optimization problem.

\subsection{Problem Formulation}

We consider a portfolio optimization problem with $n$ risky assets. 
At each rebalancing period $t$, let $\bm{x}_t$ denote the observed market features and let $\bm{r}_t \in \mathbb{R}^n$ denote the realized return vector. 
A prediction model $f_{\theta}(\cdot)$ maps the features to the predicted return vector:
\begin{equation}
    \hat{\bm{r}}_t = f_{\theta}(\bm{x}_t).
\end{equation}
Following the decision-focused learning (DFL) paradigm, the predicted returns are used as inputs to a downstream portfolio optimizer:
\begin{equation}
    \hat{\bm{w}}_t(\hat{\bm{r}}_t) 
    \in 
    \arg\max_{\bm{w} \in \mathcal{W}} 
    \Phi(\bm{w}; \hat{\bm{r}}_t),
\end{equation}
where $\hat{\bm{w}}_t \in \mathbb{R}^n$ is the portfolio induced by the predicted return vector. 
We consider the long-only budget-constrained feasible set
\begin{equation}
    \mathcal{W}
    =
    \left\{
    \bm{w} \in \mathbb{R}^n
    \mid
    \bm{1}^{\top}\bm{w}=1,\;
    \bm{w}\geq \bm{0}
    \right\}.
\end{equation}

We study two portfolio optimization formulations under this framework.

\paragraph{Return Maximization with Linear Transaction Cost.}
The first formulation considers return maximization with proportional transaction costs. 
Let $\hat{\bm{w}}_{t-1}$ denote the previous portfolio induced by past predictions and let $\kappa>0$ denote the transaction cost coefficient. 
The portfolio is obtained by solving
\begin{equation}
    \hat{\bm{w}}_t^{\mathrm{RW}}
    \in
    \arg\max_{\bm{w}\in\mathcal{W}}
    \left\{
    \hat{\bm{r}}_t^{\top}\bm{w}
    -
    \kappa \|\bm{w}-\hat{\bm{w}}_{t-1}\|_1
    \right\}.
\end{equation}
Without transaction costs, the optimizer favors assets with the highest predicted returns. 
The linear transaction cost term introduces a rebalancing threshold, so that portfolio weights are adjusted only when the predicted return advantage is sufficiently large.

\paragraph{Mean--Variance Optimization with Linear Transaction Cost.}
The second formulation adds a quadratic risk penalty. 
Given the covariance matrix $\bm{\Sigma}_t$ and risk-aversion coefficient $\lambda>0$, the portfolio is obtained by solving
\begin{equation}
    \hat{\bm{w}}_t^{\mathrm{MVO}}
    \in
    \arg\max_{\bm{w}\in\mathcal{W}}
    \left\{
    \hat{\bm{r}}_t^{\top}\bm{w}
    -
    \lambda \bm{w}^{\top}\bm{\Sigma}_t \bm{w}
    -
    \kappa \|\bm{w}-\hat{\bm{w}}_{t-1}\|_1
    \right\}.
\end{equation}
This model determines the final allocation through the trade-off among predicted return, risk penalty, and rebalancing cost.

Under the DFL framework, the predictor is trained according to the quality of its induced portfolio decisions. 
Let
\begin{equation}
    \bm{w}^*(\bm{r})
    =
    \arg\max_{\bm{w}\in\mathcal{W}}
    U_t(\bm{w};\bm{r}),
\end{equation}
where $U_t$ is the downstream MVO objective. 
The oracle and prediction-induced portfolios are
$\bm{w}_t^*=\bm{w}^*(\bm{r}_t)$ and
$\hat{\bm{w}}_t=\bm{w}^*(\hat{\bm{r}}_t)$, respectively. 
The DFL regret is
\begin{equation}
    \ell_{\mathrm{DFL}}(\hat{\bm{r}}_t,\bm{r}_t)
    =
    U_t(\bm{w}_t^*;\bm{r}_t)
    -
    U_t(\hat{\bm{w}}_t;\bm{r}_t).
\end{equation}
Since this loss is generally non-smooth, we use the SPO+ surrogate~\cite{elmachtoub2022smart}. 
The predictor therefore minimizes downstream regret rather than forecasting error.

\subsection{KKT-Based Interpretation}

We analyze the mechanism of decision-focused learning in portfolio optimization through the Karush--Kuhn--Tucker (KKT) conditions.

\paragraph{Return Maximization with Transaction Cost}Consider the return-maximization problem with linear transaction cost:
\begin{equation}
\max_{\bm w}
\quad
\hat{\bm r}^{\top}\bm w
-
\kappa \|\bm w-\bm w_{t-1}\|_1
\end{equation}
\begin{equation}
\text{s.t.}
\quad
\bm 1^{\top}\bm w = 1,
\quad
\bm w \geq \bm 0 .
\end{equation}

Since the objective is concave in $\bm w$ and the feasible set is convex, the KKT conditions characterize the global optimum.
The Lagrangian is
\begin{equation}
\mathcal{L}
=
\hat{\bm r}^{\top}\bm w
-
\kappa \|\bm w-\bm w_{t-1}\|_1
-
\nu(\bm 1^{\top}\bm w-1)
+
\bm \mu^{\top}\bm w ,
\end{equation}
where $\nu$ is the multiplier for the budget constraint and $\bm\mu \geq \bm 0$ is the multiplier for the non-negativity constraint.

Let
\begin{equation}
s_i \in \partial |w_i-w_{t-1,i}|.
\end{equation}
The stationarity condition is
\begin{equation}
\hat r_i - \kappa s_i - \nu + \mu_i = 0 .
\end{equation}

Together with primal feasibility, dual feasibility, and complementary slackness,
\begin{equation}
\bm 1^{\top}\bm w = 1,\quad
\bm w \geq \bm 0,\quad
\bm\mu \geq \bm 0,\quad
\mu_i w_i = 0,
\end{equation}
we obtain the following threshold structure.

For active assets with $w_i>0$,
\begin{equation}
\hat r_i - \kappa s_i = \nu .
\end{equation}
For inactive assets with $w_i=0$,
\begin{equation}
\hat r_i - \kappa s_i \leq \nu .
\end{equation}

Therefore, the portfolio decision is not determined by the predicted return $\hat r_i$ alone, but by the transaction-cost-adjusted marginal score
\begin{equation}
\hat r_i - \kappa s_i .
\end{equation}

\paragraph{Mean--Variance Optimization with Transaction Cost}We next extend the same analysis to the mean--variance setting with linear transaction cost:
\begin{equation}
\max_{\bm w}
\quad
\hat{\bm r}^{\top}\bm w
-
\lambda \bm w^{\top}\bm\Sigma \bm w
-
\kappa \|\bm w-\bm w_{t-1}\|_1
\end{equation}
\begin{equation}
\text{s.t.}
\quad
\bm 1^{\top}\bm w = 1,
\quad
\bm w \geq \bm 0 .
\end{equation}

The Lagrangian is
\begin{equation}
\mathcal{L}_{\mathrm{MVO}}
=
\hat{\bm r}^{\top}\bm w
-
\lambda \bm w^{\top}\bm\Sigma \bm w
-
\kappa \|\bm w-\bm w_{t-1}\|_1
-
\nu(\bm 1^{\top}\bm w-1)
+
\bm \mu^{\top}\bm w .
\end{equation}

Let
\begin{equation}
s_i \in \partial |w_i-w_{t-1,i}|.
\end{equation}
The stationarity condition is
\begin{equation}
\hat r_i
-
2\lambda(\bm\Sigma \bm w)_i
-
\kappa s_i
-
\nu
+
\mu_i
=
0 .
\end{equation}

Together with primal feasibility, dual feasibility, and complementary slackness,
\begin{equation}
\bm 1^{\top}\bm w = 1,\quad
\bm w \geq \bm 0,\quad
\bm\mu \geq \bm 0,\quad
\mu_i w_i = 0,
\end{equation}
we obtain the following threshold structure.

For active assets with $w_i>0$,
\begin{equation}
\hat r_i
-
2\lambda(\bm\Sigma \bm w)_i
-
\kappa s_i
=
\nu .
\end{equation}

For inactive assets with $w_i=0$,
\begin{equation}
\hat r_i
-
2\lambda(\bm\Sigma \bm w)_i
-
\kappa s_i
\leq
\nu .
\end{equation}

Therefore, in the mean--variance setting, the portfolio decision is governed by the risk- and transaction-cost-adjusted marginal score
\begin{equation}
\hat r_i
-
2\lambda(\bm\Sigma \bm w)_i
-
\kappa s_i .
\end{equation}

Therefore, portfolio optimization can be interpreted as a ranking process over adjusted marginal scores, rather than a direct use of predicted return levels. This perspective is consistent with prior interpretations of decision-focused learning as a ranking problem, while providing a complementary KKT-based explanation in the context of portfolio optimization with transaction costs and risk considerations.

\subsection{Empirical Observation: Prediction Inflation and Excessive Turnover}

\begin{table}[t]
\centering
\caption{
Monthly turnover (\%) of SPO+-trained MVO portfolios under different risk-aversion levels $\lambda$ across the DOW, ETF\_A, and ETF\_B datasets. 
Each value reports the mean and standard deviation over 5 random seeds. 
Increasing risk aversion does not meaningfully reduce turnover. 
Across datasets and random seeds, turnover remains unrealistically high, indicating that the instability is not simply driven by insufficient risk penalization in the MVO objective or by a particular random initialization.
}
\label{tab:turnover_lambda_risk}
\small
\setlength{\tabcolsep}{4pt}
\begin{tabular}{c|ccc}
\hline
$\lambda$ & DOW & ETF\_A & ETF\_B \\
\hline
0.1  & 94.93 $\pm$ 1.61 & 84.23 $\pm$ 6.01 & 87.04 $\pm$ 4.03 \\
1.0  & 95.49 $\pm$ 1.84 & 83.94 $\pm$ 6.11 & 86.20 $\pm$ 4.52 \\
10.0 & 95.22 $\pm$ 0.73 & 83.94 $\pm$ 5.42 & 87.37 $\pm$ 4.06 \\
20.0 & 95.51 $\pm$ 1.16 & 84.35 $\pm$ 6.28 & 87.03 $\pm$ 3.95 \\
50.0 & 95.50 $\pm$ 0.47 & 84.51 $\pm$ 5.08 & 88.00 $\pm$ 4.65 \\
\hline
\end{tabular}
\end{table}

\begin{figure*}[t]
    \centering
    \includegraphics[width=\textwidth]{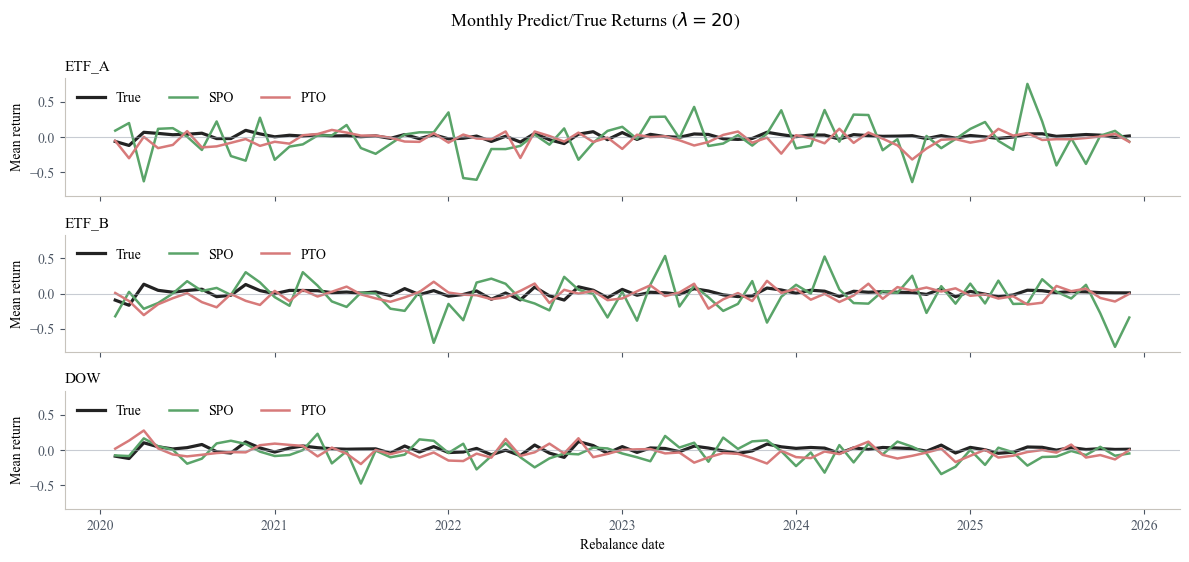}
    \caption{
    Monthly predicted and realized mean returns under $\lambda=20$ in one representative trial.  DFL predictions show much larger fluctuations than realized returns and PTO predictions, suggesting that SPO -based DFL learns decision-inducing scores rather than calibrated return forecasts.
    }
    \label{fig:monthly_pred_true_lambda20}
\end{figure*}

To examine the mechanism discussed above, we report the turnover of SPO+-trained MVO portfolios and the predicted returns at each rebalancing date. 
Table~\ref{tab:turnover_lambda_risk} summarizes monthly turnover under different risk-aversion levels over multiple random seeds, and Figure~\ref{fig:monthly_pred_true_lambda20} shows the prediction behavior of one representative trial under $\lambda=20$. 
Turnover remains high across datasets, risk-aversion levels, and random seeds, suggesting that increasing risk aversion alone cannot stabilize the portfolio.

Figure~\ref{fig:monthly_pred_true_lambda20} shows that SPO+-based DFL produces much more volatile predictions than both realized returns and PTO predictions. 
This can be understood as a mathematical consequence of SPO+-based training: the predictor enlarges return differences when doing so improves the downstream decision. 
However, such values are not suitable as realistic monthly return forecasts; they are better interpreted as decision-inducing scores rather than calibrated predictions. 
Overall, these results support our interpretation that DFL behaves as a decision-induced ranking mechanism and requires additional stabilization for practical portfolio optimization.

\section{Experiment Setup}
\subsection{Dataset}

We conduct experiments on three asset universes: DOW, ETF\_A, and ETF\_B. The DOW dataset consists of the constituents of the Dow Jones Industrial Average. ETF\_A contains eight broad-market and asset-class ETFs: EEM, EFA, JPXN, SPY, XLK, VTI, AGG, and DBC. ETF\_B contains ten sector and style ETFs: SPY, QQQ, IWM, XLK, XLF, XLE, XLV, XLY, XLP, and XLI.

Before generating return predictions, we preprocess the raw market data and construct a set of economically interpretable input features. Specifically, for each asset, we use log return, 10-day simple moving average, price bias, 14-day relative strength index, MACD difference, Bollinger-band width, volume bias, and ticker information. These features summarize recent return dynamics, trend-following signals, momentum conditions, volatility-related information, and trading-volume deviations. They are used as contextual market information for the prediction model.

The main backtesting period is from 2020 to 2026. In addition, data from 2019 are used as a warm-up period for feature construction and initial model training. The portfolio is rebalanced monthly. At each rebalancing date, we use a rolling training scheme with nine months of training data and three months of validation data for Bayesian hyperparameter optimization. After hyperparameter selection, the model generates return predictions for the next rebalancing period, and the downstream portfolio optimization problem is solved based on these predictions.

\subsection{Models and Baselines}

In this study, we consider a linear prediction model combined with a downstream mean--variance optimization (MVO) problem with transaction costs. Given the market feature vector $\bm x_t \in \mathbb{R}^d$ at rebalancing period $t$, the linear predictor generates the predicted return vector $\hat{\bm r}_t \in \mathbb{R}^n$ as
\begin{equation}
    \hat{\bm r}_t
    =
    \bm{\Theta}\bm x_t
    +
    \bm b,
\end{equation}
where $\bm{\Theta} \in \mathbb{R}^{n \times d}$ is the coefficient matrix, $\bm b \in \mathbb{R}^n$ is the intercept vector, and $n$ denotes the number of assets.

The predicted returns are then passed into the downstream portfolio optimizer. Let $\bm w_t$ denote the portfolio weight vector and $\bm w_{t-1}$ denote the portfolio held in the previous period. The downstream MVO problem is formulated as
\begin{equation}
    \bm w_t^{*}
    \in
    \arg\max_{\bm w}
    \left\{
    \hat{\bm r}_t^\top \bm w
    -
    \lambda \bm w^\top \bm \Sigma_t \bm w
    -
    \kappa \|\bm w - \bm w_{t-1}\|_1
    \right\},
\end{equation}
subject to
\begin{equation}
    \bm 1^\top \bm w = 1, 
    \qquad
    \bm w \geq \bm 0,
\end{equation}
where $\bm \Sigma_t$ is the covariance matrix estimated from historical returns, $\lambda$ controls risk aversion, and $\kappa$ controls the strength of the transaction cost penalty.

For the decision-focused learning model, the predictor is trained through the downstream optimization problem using the SPO+ surrogate loss. Since the original DFL loss is generally non-differentiable, we use the SPO+ surrogate to obtain a tractable training objective. Therefore, in the remainder of this paper, we refer to this SPO+-trained decision-focused model simply as SPO.

We compare SPO with two baseline models. The first baseline is the standard MVO strategy, which directly solves the mean--variance portfolio optimization problem using historical return estimates without training a prediction model. The second baseline is PTO-MVO, a predict-then-optimize approach. PTO-MVO uses the same linear predictor and the same downstream MVO optimizer, but the predictor is trained by minimizing prediction error rather than decision loss. Therefore, PTO-MVO evaluates whether improving return prediction accuracy leads to better downstream portfolio performance, while SPO directly optimizes the predictor according to portfolio decision quality.

\subsection{Practical Interventions for Stabilizing SPO-based Portfolios}

The empirical observations in the previous subsection suggest that SPO-based DFL may generate inflated return predictions and excessive portfolio turnover. These behaviors are closely related to the mathematical nature of SPO. Since SPO optimizes prediction models through downstream decision quality rather than pointwise prediction accuracy, the learned predictions are not necessarily calibrated return estimates in the conventional MSE sense. Instead, they may behave more like decision-inducing signals that emphasize cross-sectional ranking and portfolio selection.

However, this property does not imply that SPO is unsuitable for financial applications. On the contrary, SPO-based DFL has several attractive advantages for portfolio optimization. First, it directly aligns the learning objective with the final investment decision, rather than treating return prediction and portfolio construction as two disconnected stages. Second, compared with conventional MSE-based prediction, SPO can more naturally exploit contextual market information when such information is useful for improving downstream portfolio performance. Third, it provides a flexible framework for incorporating realistic optimization objectives and constraints into the learning process. Therefore, the purpose of this subsection is not to reject the SPO formulation, but to introduce practical interventions that make SPO-based decisions more compatible with realistic trading environments.

Simply increasing the downstream risk-aversion parameter may not be sufficient to address prediction inflation and excessive turnover. Therefore, we consider practical interventions at two stages of the SPO-based pipeline: controlling the scale of predicted returns before optimization and smoothing the realized portfolio adjustment after optimization.

\subsubsection{Prediction Scale Control}

The first intervention is to control the scale of the predicted returns before they are passed into the downstream optimizer. This treatment is motivated by a simple empirical consideration: in realistic investment environments, especially for ETFs and large-cap stocks, extremely large monthly return forecasts are difficult to justify. However, as shown in Figure~\ref{fig:monthly_pred_true_lambda20}, SPO-based DFL may produce predicted returns whose magnitudes are far beyond the realized return scale. Directly feeding such inflated predictions into the optimizer can induce overly aggressive portfolio reallocations.

Let $\hat{\bm r}_t$ denote the predicted return vector generated by the DFL model at rebalancing period $t$. In this study, we consider two simple post-processing transformations for prediction scale control: prediction clipping and min--max rescaling.

First, prediction clipping restricts each predicted return to a pre-specified realistic range:
\begin{equation}
    \tilde{\bm r}_t
    =
    \mathrm{clip}(\hat{\bm r}_t, -\gamma, \gamma),
\end{equation}
where $\gamma>0$ is a clipping threshold. The downstream optimizer then uses $\tilde{\bm r}_t$ instead of $\hat{\bm r}_t$.

This operation can be interpreted as a winsorization-type treatment. It removes extreme predicted values that fall outside the interval $[-\gamma,\gamma]$, while leaving predictions inside the interval unchanged. Therefore, prediction clipping mainly targets outliers in the predicted return scale. It does not reshape the whole prediction vector, and it preserves the original magnitude information of non-extreme predictions.

Second, min--max rescaling maps the entire prediction vector into a pre-specified symmetric return range. 
Let $c>0$ denote the half-width of the target range $[-c,c]$. 
The rescaled prediction for asset $i$ is defined as
\begin{equation}
    \tilde r_{t,i}
    =
    -c
    +
    2c
    \frac{
        \hat r_{t,i}
        -
        \min_j \hat r_{t,j}
    }{
        \max_j \hat r_{t,j}
        -
        \min_j \hat r_{t,j}
    } .
\end{equation}
In our experiments, the target range is set to a realistic monthly return scale, such as $[-0.1,0.1]$.

Unlike clipping, min--max rescaling changes the scale of the entire prediction vector within each rebalancing period. It preserves the cross-sectional ordering of predicted returns, but changes their absolute magnitudes and relative spacings. Therefore, clipping is a conservative intervention that mainly removes outliers, whereas min--max rescaling is a stronger intervention that transforms the empirical distribution of predicted returns into a controlled scale.

Both transformations do not change the SPO training objective itself. They are applied after prediction and before optimization. Their purpose is to prevent unrealistic predicted return magnitudes from directly dominating the downstream portfolio optimizer. This is particularly important when SPO-based DFL is interpreted as learning decision-relevant rankings rather than well-calibrated return levels.

\subsubsection{Partial Portfolio Adjustment}

The second intervention is to restrict the speed of portfolio adjustment. This strategy is motivated by the ranking-based interpretation of SPO. If SPO-based DFL mainly learns decision-relevant rankings rather than calibrated return levels, then the predicted returns should not necessarily be interpreted as reliable estimates of return magnitude. Instead, they can be used to identify the relative attractiveness of assets. Based on this view, we do not fully rebalance the portfolio to the optimizer's target decision. Rather, we use the target portfolio as a directional signal and adjust the current portfolio only partially.

Let $\bm w_t^{*}$ denote the target portfolio obtained from the downstream optimizer using the transformed predicted returns. We initialize the portfolio with an equally weighted allocation:
\begin{equation}
    \bm w_0
    =
    \left(
    \frac{1}{n}, \frac{1}{n}, \ldots, \frac{1}{n}
    \right)^\top,
\end{equation}
where $n$ is the number of assets. Instead of directly setting the actual portfolio to $\bm w_t^{*}$ at each rebalancing date, we update the portfolio from the previous allocation toward the target allocation:
\begin{equation}
    \bm w_t
    =
    \bm w_{t-1}
    +
    \delta
    \left(
    \bm w_t^{*} - \bm w_{t-1}
    \right),
\end{equation}
where $\delta \in (0,1]$ controls the adjustment speed. When $\delta=1$, the portfolio is fully rebalanced to the target portfolio. When $\delta<1$, only a fraction of the gap between the current portfolio and the target portfolio is closed at each rebalancing date.

This strategy directly reduces turnover by smoothing the realized portfolio path over time. More importantly, it is consistent with the interpretation that SPO provides a ranking-oriented decision signal. The optimizer's output indicates the direction in which the portfolio should move, but the actual trading decision is adjusted gradually from the existing allocation to avoid excessive reallocation. Starting from an equally weighted portfolio also provides a neutral initial allocation before the SPO-based signal is gradually incorporated.

This adjustment rule is closer to practical portfolio management, where investors rarely replace the entire portfolio at every rebalancing date. In real markets, transaction costs, liquidity constraints, market impact, and execution risks all discourage abrupt full rebalancing. Therefore, partial portfolio adjustment serves as a simple and practical mechanism for translating the SPO-induced target portfolio into a smoother realized trading strategy.

Overall, prediction scale control and partial portfolio adjustment intervene at different stages of the SPO-based pipeline. Prediction clipping removes extreme predicted values while preserving non-extreme magnitude information. Min--max rescaling maps the entire prediction vector into a realistic return scale and preserves the cross-sectional ranking signal. Partial portfolio adjustment then stabilizes the realized portfolio decision after optimization by gradually moving from the current allocation toward the target allocation. Together, these interventions address unrealistic prediction magnitudes and excessive turnover without changing the SPO training objective itself.

\subsection{Experimental Configuration}

Table~\ref{tab:experimental_config} summarizes the main experimental settings. 
The experiment matrix consists of six SPO variants, three asset universes, and five risk-aversion levels.

\begin{table}[t]
\centering
\caption{Summary of experimental configuration.}
\label{tab:experimental_config}
\small
\begin{tabular}{ll}
\toprule
Item & Setting \\
\midrule
Variants 
& \begin{tabular}[c]{@{}l@{}}
Standard, Clip, Rescale, Adj, \\
Clip+Adj, Rescale+Adj
\end{tabular} \\
Datasets 
& DOW, ETF\_A, ETF\_B \\
Risk aversion 
& $\lambda \in \{0.1,1.0,10.0,20.0,50.0\}$ \\
Fee rate
& $\kappa=0.005$\\
Backtest period 
& 2020--2026 \\
Warm-up period 
& 2019 \\
Rebalancing 
& Monthly \\
Rolling window 
& 9 months training + 3 months validation \\
Prediction model 
& Linear predictor trained with SPO+ \\
Optimizer 
& MVO with $\ell_1$ transaction cost \\
Clip 
& $\gamma=0.1$ \\
Rescale 
& Min--max normalization to $[-0.1,0.1]$ \\
Adjustment 
& $\delta=0.1$ \\
Covariance window 
& 220 trading days \\
Covariance regularization 
& $10^{-6}$ \\
Baselines 
& PTO-MVO, MVO \\
\bottomrule
\end{tabular}
\end{table}

\section{Results and Discussion}
\begin{table*}[t]
\centering
\caption{
Performance comparison under different values of $\lambda$. 
}
\label{tab:ablation_full}
\scriptsize
\resizebox{\textwidth}{!}{
\begin{tabular}{c|c|ccccc|ccccc|ccccc}
\hline
$\lambda$ & Setting 
& \multicolumn{5}{c|}{DOW} 
& \multicolumn{5}{c|}{ETF\_A} 
& \multicolumn{5}{c}{ETF\_B} \\
\cline{3-17}
& & Ret. & Vol. & TO & MDD & SR 
& Ret. & Vol. & TO & MDD & SR 
& Ret. & Vol. & TO & MDD & SR \\
\hline

\multirow{8}{*}{0.1}
& Standard & -0.0032 & 0.3509 & 0.9571 & -0.6352 & -0.0092 & 0.1111 & 0.2330 & 0.8429 & -0.2767 & 0.4769 & 0.0137 & 0.2461 & 0.9000 & -0.3790 & 0.0555 \\
& Clip & 0.0163 & 0.1808 & 0.5948 & \textbf{-0.2794} & 0.0899 & 0.0910 & 0.1973 & 0.5580 & -0.3299 & 0.4613 & 0.0159 & 0.2025 & 0.5548 & \textbf{-0.3486} & 0.0784 \\
& Rescale & -0.0452 & 0.3504 & 0.9571 & -0.6631 & -0.1291 & 0.0749 & 0.2332 & 0.8286 & -0.2908 & 0.3213 & -0.0601 & 0.2782 & 0.8857 & -0.5643 & -0.2158 \\
& Adj & 0.1387 & 0.2178 & 0.0963 & -0.3365 & 0.6366 & 0.1345 & 0.1721 & 0.0861 & \textbf{-0.2478} & 0.7816 & 0.1232 & 0.2115 & 0.0900 & -0.3818 & 0.5823 \\
& Clip+Adj & 0.0875 & \textbf{0.1782} & \textbf{0.0578} & -0.3182 & 0.4907 & 0.1056 & \textbf{0.1527} & \textbf{0.0573} & -0.2575 & 0.6917 & 0.1075 & \textbf{0.1924} & \textbf{0.0565} & -0.3608 & 0.5588 \\
& Rescale+Adj & \textbf{0.1726} & 0.2171 & 0.0948 & -0.3368 & \textbf{0.7951} & \textbf{0.1467} & 0.1728 & 0.0845 & -0.2539 & \textbf{0.8492} & \textbf{0.1404} & 0.2122 & 0.0881 & -0.3862 & \textbf{0.6616} \\
& PTO & 0.0306 & 0.2947 & 0.9296 & -0.6248 & 0.1039 & -0.0410 & 0.2046 & 0.8310 & -0.4113 & -0.2002 & 0.0929 & 0.2422 & 0.9014 & -0.3443 & 0.3834 \\
& MVO & 0.2497 & 0.4597 & 0.2224 & -0.5926 & 0.5431 & 0.1622 & 0.2479 & 0.1938 & -0.3115 & 0.6545 & 0.1636 & 0.2763 & 0.2022 & -0.3115 & 0.5920 \\
\hline

\multirow{8}{*}{1.0}
& Standard & 0.0259 & 0.3485 & 0.9571 & -0.6352 & 0.0742 & 0.1123 & 0.2332 & 0.8429 & -0.2767 & 0.4816 & 0.0137 & 0.2461 & 0.9000 & -0.3790 & 0.0555 \\
& Clip & 0.0086 & 0.1836 & 0.6066 & \textbf{-0.2836} & 0.0468 & 0.0898 & 0.1973 & 0.5573 & -0.3299 & 0.4553 & 0.0050 & 0.2030 & 0.5625 & -0.3836 & 0.0245 \\
& Rescale & -0.0204 & 0.3482 & 0.9571 & -0.6631 & -0.0587 & 0.0749 & 0.2332 & 0.8286 & -0.2908 & 0.3213 & -0.0601 & 0.2782 & 0.8857 & -0.5643 & -0.2158 \\
& Adj & 0.1320 & 0.2171 & 0.0960 & -0.3365 & 0.6079 & 0.1344 & 0.1719 & 0.0861 & \textbf{-0.2478} & 0.7819 & 0.1239 & 0.2117 & 0.0896 & -0.3818 & 0.5850 \\
& Clip+Adj & 0.0849 & \textbf{0.1785} & \textbf{0.0577} & -0.3187 & 0.4758 & 0.1054 & \textbf{0.1529} & \textbf{0.0571} & -0.2575 & 0.6890 & 0.1055 & \textbf{0.1934} & \textbf{0.0560} & \textbf{-0.3608} & 0.5457 \\
& Rescale+Adj & \textbf{0.1643} & 0.2161 & 0.0950 & -0.3370 & \textbf{0.7602} & \textbf{0.1463} & 0.1726 & 0.0844 & -0.2539 & \textbf{0.8477} & \textbf{0.1402} & 0.2120 & 0.0883 & -0.3862 & \textbf{0.6615} \\
& PTO & 0.0306 & 0.2947 & 0.9296 & -0.6248 & 0.1039 & -0.0410 & 0.2046 & 0.8310 & -0.4113 & -0.2002 & 0.0929 & 0.2422 & 0.9014 & -0.3443 & 0.3834 \\
& MVO & 0.2227 & 0.3975 & 0.2625 & -0.4548 & 0.5603 & 0.1039 & 0.2360 & 0.2481 & -0.3115 & 0.4404 & 0.1575 & 0.2701 & 0.2305 & -0.3115 & 0.5830 \\
\hline

\multirow{8}{*}{10.0}
& Standard & -0.0130 & 0.3368 & 0.9571 & -0.6096 & -0.0387 & 0.0994 & 0.2333 & 0.8714 & -0.2996 & 0.4262 & 0.0671 & 0.2490 & 0.9143 & -0.3790 & 0.2693 \\
& Clip & 0.0053 & 0.1830 & 0.5886 & \textbf{-0.2837} & 0.0290 & 0.0798 & 0.1931 & 0.5534 & -0.3182 & 0.4132 & 0.0146 & 0.2013 & 0.5561 & \textbf{-0.3400} & 0.0728 \\
& Rescale & -0.0258 & 0.3189 & 0.9553 & -0.6403 & -0.0808 & 0.0499 & 0.2297 & 0.8329 & -0.3107 & 0.2174 & 0.0035 & 0.2685 & 0.8984 & -0.5230 & 0.0132 \\
& Adj & 0.1221 & 0.2121 & 0.0962 & -0.3365 & 0.5755 & 0.1328 & 0.1723 & 0.0868 & \textbf{-0.2478} & 0.7703 & 0.1295 & 0.2118 & 0.0903 & -0.3818 & 0.6115 \\
& Clip+Adj & 0.0900 & \textbf{0.1790} & \textbf{0.0576} & -0.3186 & 0.5027 & 0.1045 & \textbf{0.1514} & \textbf{0.0557} & -0.2575 & 0.6904 & 0.1069 & \textbf{0.1912} & \textbf{0.0554} & -0.3608 & 0.5592 \\
& Rescale+Adj & \textbf{0.1597} & 0.2079 & 0.0939 & -0.3355 & \textbf{0.7682} & \textbf{0.1464} & 0.1717 & 0.0843 & -0.2539 & \textbf{0.8525} & \textbf{0.1437} & 0.2120 & 0.0866 & -0.3862 & \textbf{0.6780} \\
& PTO & 0.0306 & 0.2947 & 0.9296 & -0.6248 & 0.1039 & -0.0410 & 0.2046 & 0.8310 & -0.4113 & -0.2002 & 0.0971 & 0.2423 & 0.8956 & -0.3443 & 0.4005 \\
& MVO & 0.0526 & 0.2034 & 0.2892 & -0.3024 & 0.2587 & 0.0307 & 0.1078 & 0.1421 & -0.2126 & 0.2846 & 0.0493 & 0.1788 & 0.2480 & -0.2695 & 0.2756 \\
\hline

\multirow{8}{*}{20.0}
& Standard & 0.0034 & 0.3429 & 0.9581 & -0.6333 & 0.0100 & 0.1006 & 0.2289 & 0.8714 & -0.2954 & 0.4394 & 0.0263 & 0.2803 & 0.9000 & -0.5643 & 0.0939 \\
& Clip & -0.0039 & 0.1826 & 0.5699 & \textbf{-0.2908} & -0.0214 & 0.0725 & 0.1892 & 0.5899 & -0.3299 & 0.3834 & 0.0114 & 0.2015 & 0.5554 & \textbf{-0.3363} & 0.0566 \\
& Rescale & -0.0269 & 0.3046 & 0.9492 & -0.6098 & -0.0884 & 0.0487 & 0.2277 & 0.8291 & -0.3036 & 0.2138 & -0.0108 & 0.2781 & 0.8942 & -0.5643 & -0.0387 \\
& Adj & 0.1252 & 0.2146 & 0.0962 & -0.3365 & 0.5835 & 0.1334 & 0.1703 & 0.0869 & \textbf{-0.2478} & 0.7834 & 0.1240 & 0.2164 & 0.0902 & -0.4027 & 0.5730 \\
& Clip+Adj & 0.0854 & \textbf{0.1786} & \textbf{0.0575} & -0.3186 & 0.4779 & 0.1054 & \textbf{0.1513} & \textbf{0.0570} & -0.2575 & 0.6964 & 0.1037 & \textbf{0.1924} & \textbf{0.0554} & -0.3608 & 0.5388 \\
& Rescale+Adj & \textbf{0.1417} & 0.2050 & 0.0934 & -0.3369 & \textbf{0.6916} & \textbf{0.1472} & 0.1696 & 0.0841 & -0.2514 & \textbf{0.8675} & \textbf{0.1366} & 0.2157 & 0.0874 & -0.4031 & \textbf{0.6331} \\
& PTO & 0.0258 & 0.2932 & 0.9341 & -0.6245 & 0.0880 & -0.0410 & 0.2046 & 0.8310 & -0.4113 & -0.2002 & 0.0968 & 0.2423 & 0.8968 & -0.3443 & 0.3996 \\
& MVO & 0.0375 & 0.1763 & 0.2450 & -0.2668 & 0.2128 & 0.0194 & 0.0785 & 0.0907 & -0.1698 & 0.2475 & 0.0522 & 0.1651 & 0.1580 & -0.2545 & 0.3164 \\
\hline

\multirow{8}{*}{50.0}
& Standard & 0.0328 & 0.3234 & 0.9513 & -0.6177 & 0.1015 & 0.0374 & 0.2236 & 0.8584 & -0.3373 & 0.1671 & 0.0358 & 0.2448 & 0.8932 & -0.3790 & 0.1463 \\
& Clip & 0.0103 & 0.1801 & 0.5735 & \textbf{-0.2825} & 0.0572 & 0.0751 & 0.1901 & 0.5616 & -0.3276 & 0.3949 & 0.0340 & 0.1927 & 0.5773 & \textbf{-0.3245} & 0.1763 \\
& Rescale & -0.0266 & 0.2701 & 0.9382 & -0.5748 & -0.0985 & 0.0083 & 0.2217 & 0.8352 & -0.3343 & 0.0375 & 0.0054 & 0.2369 & 0.8984 & -0.3850 & 0.0228 \\
& Adj & 0.1126 & 0.2067 & 0.0963 & -0.3365 & 0.5448 & 0.1282 & 0.1747 & 0.0856 & -0.2719 & 0.7337 & 0.1283 & 0.2112 & 0.0898 & -0.3818 & 0.6076 \\
& Clip+Adj & 0.0812 & \textbf{0.1778} & \textbf{0.0572} & -0.3177 & 0.4564 & 0.1049 & \textbf{0.1509} & \textbf{0.0560} & \textbf{-0.2575} & 0.6950 & 0.0970 & \textbf{0.1909} & \textbf{0.0553} & -0.3608 & 0.5083 \\
& Rescale+Adj & \textbf{0.1194} & 0.1940 & 0.0928 & -0.3354 & \textbf{0.6153} & \textbf{0.1422} & 0.1717 & 0.0833 & -0.2722 & \textbf{0.8284} & \textbf{0.1424} & 0.2081 & 0.0853 & -0.3773 & \textbf{0.6846} \\
& PTO & 0.0288 & 0.2910 & 0.9432 & -0.6076 & 0.0988 & -0.0330 & 0.2031 & 0.8376 & -0.4113 & -0.1627 & 0.0963 & 0.2418 & 0.9018 & -0.3443 & 0.3982 \\
& MVO & 0.0374 & 0.1593 & 0.1757 & -0.2375 & 0.2348 & 0.0153 & 0.0659 & 0.0446 & -0.1558 & 0.2330 & 0.0542 & 0.1606 & 0.0918 & -0.2610 & 0.3373 \\
\hline
\end{tabular}
}
\par\smallskip
\begin{minipage}{\textwidth}
\footnotesize
\textit{Note.} Ret. and Vol. are annualized. 
TO denotes monthly turnover reported as a proportion. 
MDD is computed over the full backtesting period. 
Bold values indicate the best metric among the six SPO variants 
(Standard, Clip, Rescale, Adj, Clip+Adj, Rescale+Adj). 
Ret., SR, and MDD are maximized, while Vol. and TO are minimized. 
PTO and MVO are reported as baseline references and are not included in the bold comparison.
\end{minipage}
\end{table*}

\paragraph{Ablation Experiment}
Table~\ref{tab:ablation_full} reports the ablation results under different values of the risk-aversion parameter $\lambda$. 
Overall, the results show that portfolio adjustment plays the most direct role in reducing turnover. 
The Standard variant exhibits very high turnover across datasets, indicating that directly applying the SPO-trained predictor to the downstream optimizer can lead to unstable portfolio reallocations. 
Rescale also maintains high turnover when used alone, even though the predicted returns are mapped into the same bounded range as the clipping operation. 
This is because Rescale preserves the cross-sectional ranking of the prediction vector within each rebalancing period, so the downstream optimizer may still react to the same ranking signal as in the Standard setting.

Clipping alone reduces turnover and volatility compared with Standard in many cases. 
This suggests that suppressing extreme predicted values can weaken overly dominant prediction signals before they enter the optimizer. 
However, clipping does not directly constrain the realized portfolio path, and therefore its ability to control turnover remains limited when it is used without portfolio adjustment.

By contrast, Adj, Clip+Adj, and Rescale+Adj substantially reduce turnover across all datasets and values of $\lambda$. 
This confirms that partial portfolio adjustment is the main mechanism for smoothing allocation changes. 
Among these stabilized variants, Clip+Adj achieves the lowest turnover and the lowest or near-lowest volatility in most cases, making it the most conservative and stable strategy. 
At the same time, Rescale+Adj often achieves higher returns and Sharpe ratios, suggesting that preserving the ranking signal while smoothing portfolio transitions can produce a more aggressive but potentially more profitable strategy.

These findings are consistent with the decision-induced ranking interpretation of SPO-based portfolio optimization. 
The instability is not merely caused by the absolute scale of predicted returns, because Rescale changes the numerical range but preserves the asset ranking and therefore leaves much of the decision-inducing signal unchanged. 
Rather, the downstream optimizer is sensitive to both the ranking and the relative gaps among predicted scores. 
Therefore, practical stabilization requires not only controlling unrealistic prediction magnitudes, but also smoothing how ranking-driven signals are translated into actual portfolio trades. 
Overall, the results suggest that turnover control at the portfolio level and output constraints based on realistic financial considerations are effective for stabilizing SPO-based DFL models. 
They also indicate that other practically motivated correction mechanisms may be worth exploring for making DFL-based portfolio strategies more implementable in real financial settings.

\paragraph{Predicted Return Distribution}

\begin{figure*}[t]
    \centering
    \includegraphics[width=0.95\textwidth]{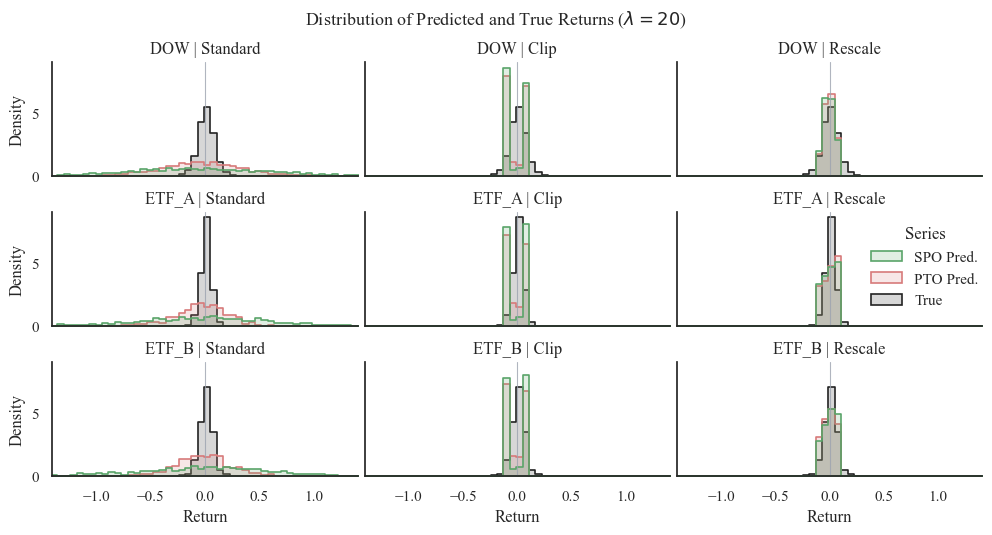}
    \caption{
    Distribution of predicted and realized returns under $\lambda=20$ for Standard, Clip, and Rescale settings. 
    The Standard setting produces widely dispersed SPO predictions relative to realized returns and PTO predictions. 
    Clipping suppresses extreme predicted values and concentrates predictions around a realistic return range. 
    Rescale maps predictions into a bounded range through min--max normalization, but preserves the cross-sectional ranking structure within each rebalancing period. 
    The comparison suggests that prediction magnitude, ranking structure, and score gaps jointly affect downstream portfolio decisions.
    }
    \label{fig:return_distribution}
\end{figure*}

Figure~\ref{fig:return_distribution} compares the distributions of predicted and realized returns under $\lambda=20$ for Standard, Clip, and Rescale settings. 
Under the Standard setting, the SPO predictions are much more widely dispersed than both the realized returns and PTO predictions across the three datasets. 
This confirms that the SPO-trained model does not primarily produce calibrated return forecasts, but instead tends to amplify prediction magnitudes in order to induce clearer downstream portfolio decisions.

After clipping, the predicted return distributions become much more concentrated around a realistic return range. 
This suggests that clipping effectively suppresses extreme predicted values before they enter the downstream optimizer. 
As a result, the optimizer is less exposed to overly large prediction gaps, which helps explain the lower volatility and turnover observed for Clip and Clip+Adj in Table~\ref{tab:ablation_full}.

By contrast, Rescale also maps predictions into a bounded numerical range, but it changes the entire prediction vector through min--max normalization while preserving the cross-sectional ranking within each rebalancing period. 
The resulting distributions are closer to the realized return scale than those under Standard, but this transformation does not necessarily weaken the ranking signal used by the optimizer. 
This explains why Rescale alone does not consistently reduce turnover in Table~\ref{tab:ablation_full}: the numerical scale is controlled, but the relative ordering that drives the portfolio decision remains largely unchanged.

Overall, the distributional evidence supports the interpretation that SPO-based portfolio decisions are affected not only by the absolute magnitude of predicted returns, but also by the ranking structure and score gaps among assets. 
Therefore, prediction-level transformations such as clipping and rescaling should be understood as mechanisms for controlling the decision signal before optimization, rather than as conventional forecast calibration methods.

\section{Conclusion}  
This study investigated the behavior of SPO-based decision-focused learning in portfolio optimization, with particular attention to prediction inflation, ranking-driven allocation, and excessive turnover.  Although DFL aims to align the training objective with downstream decision quality, our analysis suggests that, in portfolio optimization, this alignment can lead the predictor to generate decision-inducing signals rather than well-calibrated return forecasts.  Through a KKT-based interpretation, we showed that portfolio decisions can be viewed as a ranking process over adjusted marginal scores that incorporate predicted returns, risk penalties, and transaction costs.  This provides a complementary explanation for why SPO-based learning may emphasize relative asset ordering and produce aggressive portfolio reallocations.  Empirically, we observed that SPO-trained models can generate inflated predicted returns and high turnover under standard mean--variance optimization with transaction costs.  Increasing the risk-aversion parameter alone did not effectively eliminate this instability, suggesting that the problem is not merely caused by insufficient risk penalization in the downstream optimizer.  The ablation results further showed that prediction-level transformations and portfolio-level adjustment play different roles.  Clipping suppresses extreme predicted values, while Rescale controls the numerical range but preserves cross-sectional ranking.  Partial portfolio adjustment directly smooths allocation changes and substantially reduces turnover across datasets and risk-aversion levels.  Among the examined variants, Clip+Adj provides the most conservative and stable turnover control, while Rescale+Adj often achieves stronger returns and Sharpe ratios by preserving more of the SPO-induced ranking signal.  These results suggest a practical trade-off between exploiting decision-focused signals and maintaining implementable trading behavior.  Overall, the findings indicate that realistic output constraints and portfolio-level turnover control are effective tools for stabilizing SPO-based portfolio strategies. 

\section{Future Work}

Several directions remain for future work. First, it would be valuable to examine whether the high turnover and prediction-scale inflation observed in this study persist under different penalty formulations and predictive models. Transaction-cost penalties, turnover penalties, and other regularization terms may reshape the decision signal induced by SPO-type objectives, while different forecasting models may affect the scale, stability, and ranking structure of predicted returns. Such analysis would help clarify whether these behaviors are intrinsic to the SPO-based portfolio learning framework or are induced or mitigated by specific modeling and regularization choices.

Second, future work may examine whether practical portfolio-improvement mechanisms should be applied after prediction or incorporated directly into the training process. Although embedding transaction-cost control, turnover regularization, or partial adjustment rules into the decision-focused objective may better align learning with implementable portfolio decisions, it may also introduce additional modeling complexity and reduce interpretability. Comparing post-processing approaches with end-to-end training formulations would provide further insight into how trading frictions reshape the learning signal.

\bibliographystyle{ACM-Reference-Format}
\bibliography{ref}
\end{document}